\documentclass[fleqn,usenatbib]{mnras}

\usepackage{newtxtext,newtxmath}

\usepackage[T1]{fontenc}
\usepackage{relsize}
\usepackage{subfig}
\usepackage{hyperref}

\DeclareRobustCommand{\VAN}[3]{#2}
\let\VANthebibliography\thebibliography
\def\thebibliography{\DeclareRobustCommand{\VAN}[3]{##3}\VANthebibliography}

\usepackage{graphicx}	
\usepackage{amsmath}	

\newcommand{\be}{\begin{equation}}
\newcommand{\ee}{\end{equation}}

\title[Mountains from lattice pressure]{Crustal lattice pressure as a source of neutron star mountains}

\author[D. I. Jones \& T. J. Hutchins]{
D. I. Jones,$^{1}$\thanks{E-mail: d.i.jones@soton.ac.uk}
T. J. Hutchins,$^{1}$
\\
$^{1}$Mathematical Sciences and STAG Research Centre, University of Southampton, Southampton SO17 1BJ, United Kingdom
}

\date{Accepted XXX. Received YYY; in original form ZZZ}

\pubyear{2023}

\begin{document}
\label{firstpage}
\pagerange{\pageref{firstpage}--\pageref{lastpage}}
\maketitle

\begin{abstract}

The spin frequencies of neutron stars in low-mass X-ray binaries may be limited by the emission of gravitational waves.  A candidate for producing such steady emission is a mass asymmetry, or ``mountain", sourced by temperature asymmetries in the star's crust.  A number of studies have examined temperature-induced shifts in the crustal capture layers between one nuclear species and another to produce this asymmetry, with the presence of capture layers in the deep crust being needed to produce the required mass asymmetries.  However, modern equation of state calculations cast doubt on the existence of such deep capture layers.  Motivated by this, we investigated an alternative source of temperature dependence in the equation of state, coming from the pressure supplied by the solid crustal lattice itself.  We show that temperature-induced perturbations in this pressure, while small, may be significant.  We therefore advocate for more detailed calculations, self-consistently calculating both the temperature asymmetries, the perturbations in crustal lattice pressure, and the consequent mass asymmetries, to establish if this is a viable mechanism for explaining the observed distribution of low-mass X-ray binary spin frequencies.  Furthermore, the crustal lattice pressure mechanism does not require accretion, extending the possibility for such thermoelastic mountains to include both accreting and isolated neutron stars.

\end{abstract}


\section{Introduction}  \label{sect:introduction}

Spinning non-axisymmetric neutron stars (NSs) are a potential source of detectable continuous gravitational waves (CGWs); see \citet{rj_24} for a recent review.  In order to support a non-axisymmetric shape, some deforming force beyond the axisymmetric centrifugal one must act, so that the star has what is colloquially referred to as a \emph{mountain}.  

The two types of deforming force to consider are magnetic and elastic ones; see \citet{gg_18} for a review.  In the case of \emph{magnetic mountains}, the Lorentz forces due to a non-axisymmetric magnetic field support the mass distortion.  The existence of magnetic fields in NSs seems clear, with the (external) magnetic field having  been invoked since the discovery of the first pulsars to explain their pulsation mechanism and their gradual spin-down.

In the case of \emph{elastic mountains}, it is elastic strains in the solid crust that are responsible for the deformations. There have been several studies of the \emph{maximum} mountains that a neutron star's crust can support, limited by the finite shear modulus and breaking strain of the crustal material \citep{ucb_00, Owen_2005, Haskell_2006, Johnson-McDaniel_2013, Gittins_2020, Gittins_2021, Morales_2022}. However, in the case of elastic mountains, the \emph{cause} of the non-axisymmetry is less clear.  The changing centrifugal forces in a spinning-up or spinning-down star are necessarily axisymmetric, so cannot directly contribute to CGW emission.  It is possible that crustal fractures in response to spin-up or spin-down may result in non-axisymmetric distortions \citep{fhh_18},  but the few detailed analyses of mountain formation following crustal fracture concern themselves with \emph{maximum} rather than \emph{likely} mountain sizes
\citep{gc_22, gj_24}.

A possible cause of elastic distortions was proposed in \citet{bild_98}, and developed in \citet{ucb_00}, who noted that in an accreting neutron star in a low-mass X-ray binary (LMXB), as an element of nuclear matter is pushed deeper into the crust,  the transition from one nuclear species to another is slightly temperature dependent.  This gives the crustal equation of state (EoS) an effective temperature-dependence, so that a temperature asymmetry translates into a mass density asymmetry.  \citet{ucb_00} (hereafter UCB) noted that asymmetries in the nuclear burning rate and/or the nuclear charge-to/mass ratio could seed the temperature asymmetries.  They did not, however, attempt a first principles calculation of the formation of such asymmetries in the burning rate or composition.

In an attempt to supply the required temperature asymmetry, \citet{oj_20} made use of the fact that accreting NSs in LMXBs have (albeit weak) magnetic fields.  Such fields make the thermal conductivity tensor anisotropic, with heat being conducted more easily along the magnetic field lines than perpendicular to the field lines.  This anisotropy in thermal conductivity naturally produces a temperature asymmetry.  The crust-only model of \cite{oj_20} was recently extended to the entire star by \citet{hj_23}, who found somewhat larger temperature perturbations for a given field strength.    

In this paper, we seek to enlarge the class of NS candidates for what we might term \emph{thermoelastic mountains} beyond NSs in LMXBs by considering different sources of temperature-dependence in the crustal EoS.  In fact, we simply revisit several thermal contributions to dense matter that are relatively well studied, but have not been applied to the CGW problem thus far.

In Section \ref{sect:capture_layers} we summarise the shifting capture layer model of \cite{bild_98}, and critique it in the light of recent crustal EoS calculations.  In Section \ref{sect:electrons} we consider the thermal contributions to the relativistic Fermi electron gas that provides most of the pressure in the outer crust, while in Section \ref{sect:neutrons} we consider the thermal contributions from the non-relativistic neutron gas that provides most of the pressure in the inner crust.  As we discuss below, neither of these seem well suited to sustaining thermoelastic mountains, as they are perturbations in fluid components.  For this reason, in Section \ref{sect:lattice} we consider the thermal contribution to the crustal lattice pressure.  In Section \ref{sect:comparison} we briefly compare the relative sizes of the pressure perturbations from lattice pressure to those considered in UCB.  In Section \ref{sect:discussion} we summarise our findings, and comment on possible avenues for future research.

When parameterising formulae, we use subscripts to denote the power of ten of the cgs unit to which we normalise a quantity, e.g.\ for density, $\rho_{12} \equiv \rho / 10^{12}$\,g\,cm$^{-3}$.

\section{Thermal Mountains from capture layer shifts} \label{sect:capture_layers}

In \citet{bild_98} it was suggested that GW torques from a sufficiently large mountain could be limiting the spin period of accreting NSs, with angular momentum transferred from the accretion disk being balanced by the loss of angular momentum due to gravitational radiation.

If one assumes that \textit{all} of the spin-down energy from the NS is radiated away as GWs, the mass quadrupole $\Tilde{Q}_{22}$ which balances the accretion torque is (equation (1) of \citealp{bild_98})
\begin{equation}\label{eq: Torque Balance Quadrupole}
  \Tilde{Q}_{22} = 4.5 \times 10^{37} \text{g cm}^{2} \, \bigg( \frac{\langle \Dot{M} \rangle}{10^{-9} \, M_{\odot}  \text{yr}^{-1}} \biggr)^{1/2} \, \bigg( \frac{300 \, \text{Hz}}{\nu} \biggr)^{5/2} \, ,
\end{equation}
where $\langle \Dot{M} \rangle$ is a \textit{time-averaged} mass accretion rate and $\nu$ is the spin frequency of the star. Mass accretion rates of neutron stars in LMXBs are typically in the region $10^{-11} \lesssim \dot{M} \lesssim 10^{-8}$ M$_{\odot}$ yr$^{-1}$, with spin frequencies in the region $200 \lesssim \nu \lesssim 700$ Hz. Such considerations imply that $\Tilde{Q}_{22} \sim 10^{36-39}$ g cm$^{2}$ in order to attain torque balance for most systems if gravitational radiation is the only source of spin-down energy loss.  Bildsten suggested that such a quadrupole could be built by thermally-induced shifts in electron capture layers.

\subsection{Mass quadrupole moment from capture layer shifts}

UCB built a detailed model of the capture shifts. From their Eq. (55), the fiducial quadrupole moment generated by an individual capture layer in the outer crust (i.e. before the neutron drip point) can be approximated as 



\begin{equation}\label{eq: UCB Fiducial Q_22}
  Q_{\text{fid}} \approx 1.3 \times 10^{36} \, \text{g cm}^{2} \, R^4_6 \, \bigg( \frac{T}{10^8 \, \text{K}} \biggr) \, \bigg( \frac{\delta T_{22} /T }{1\%} \biggr) \, \bigg( \frac{E_{\text{cap}}}{30 \, \text{MeV}} \biggr)^3 \, ,
\end{equation}
where $E_{\text{cap}}$ is the threshold energy, equivalent to the electron chemical potential $\mu_e$ at the transition between one nuclear species and the next.  The above formula implies that, for a canonical $10$ km neutron star  with a crustal temperature $\sim 10^8$ K and quadrupolar ($l=m=2$) temperature asymmetry $\delta T_{22}/T \sim 1\%$, a single electron capture layer with threshold energy $E_{\text{cap}} \sim 30$ MeV produces a mass quadrupole which can be up to 3 orders of magnitude smaller than what \citet{bild_98} estimated would be required to establish torque balance, depending on the accretion rate and spin period of the system (see below).

In their calculation, UCB assumed (partially) the composition of the accreted crust to follow the series of non-equilibrium reactions listed in Tables 1 and 2 of \citet{hz_90_first} (the values of $\mu_e$ at each capture layer being listed in the sixth column of Table 2 of \citealt{hz_90_second}). The maximum value of the electron chemical potential (and therefore threshold energy) calculated by \citet{hz_90_first, hz_90_second} (denoted collectively HZ90 hereafter) was $\mu_e \equiv E_{\text{max}}^{\text{HZ90}} = 43.69$ MeV, suggesting that none of these capture layers are, individually, likely capable of producing a $Q_{22}$ large enough to attain torque balance as determined by equation (\ref{eq: Torque Balance Quadrupole}).

UCB did, however, note that the \textit{total} mass quadrupole of the NS $Q_{\text{tot}}$ is (approximately) a linear sum of the quadrupole moments generated in each capture layer individually. Combining the available data for $E_{\text{cap}}$ given in HZ90 and Eq. \eqref{eq: UCB Fiducial Q_22}, one finds that the total mass quadrupole assuming a temperature asymmetry $\delta T_{22} / T \sim 1\%$ is 
\begin{equation}\label{eq: HZ Capture Layers}
    Q_{\text{tot}} \approx 1.3 \times 10^{36} \, \text{g cm}^{2} \, 
    \mathlarger{\sum}^{19}_i \bigg[ \frac{E_{\text{cap}}^i}{30 \, \text{MeV}} \biggr]^3 \approx 4 \times 10^{37} \, \text{g cm}^{2} \, ,
\end{equation}
where the summation includes all 19 capture layers listed in Table 2 of \citet{hz_90_second}. Due to the strong scaling with the threshold energy (which increases monotonically through the crust), contributions from deep capture layers can be up to three orders of magnitude larger than those from shallow layers, and thus it is the deeper layers which dominate the total mass quadrupole.

One should bear in mind however that this calculation is only an approximation, for three reasons. Firstly, Eq. \eqref{eq: UCB Fiducial Q_22} is, strictly speaking, only valid in the outer crust. For simplicity, we summed over a number of capture layers that are present in the inner crust to obtain the result \eqref{eq: HZ Capture Layers}, since a full calculation requires numerically integrating over any geometrically `thick' capture layers (\textit{cf}. Eq. 56 in UCB and surrounding text). Secondly, as noted in UCB, shallow capture layers can produce \textit{negative} quadrupole moments, in which case some capture layers may even cancel each other out. Thirdly, Eq. \eqref{eq: UCB Fiducial Q_22} neglects to quantify the elastic response of the crust to changes in composition. 

In their detailed numerical scheme, UCB identified that actually `the crust prefers to sink in response to the shift in capture layers', rather than spreading out laterally. This `sinking penalty' reduces the actual mass quadrupole significantly below that of the fiducial estimate \eqref{eq: UCB Fiducial Q_22} (see their Fig. 15), by up to a factor of 10 for deep capture layers, and as much as a factor of 50 for shallow layers.



Even without considering elastic effects, the above result suggests that capture layer shifts are only capable of determining the spin equilibrium of LMXBs which are spinning very fast ($\sim 600$ Hz). Also, as shown in \citet{oj_20, hj_23} (and UCB to a lesser extent), temperature asymmetries close to the (assumed) percent level are only likely to persist in systems which are strongly accreting.  On this basis, it is therefore not clear that sufficiently large quadrupoles can be generated in the full set of LMXBs.

When extrapolating to higher values of the threshold energy however, UCB showed that a \textit{single} additional `artificial' capture layer (i.e. with $E_{\text{cap}} > 44$ MeV) added ad-hoc near the bottom of the crust could generate mass quadrupoles in excess of $10^{37}$ g cm$^{2}$, but only if the threshold energy exceeds $ \sim 80$ MeV.  UCB then considered the possibility of there being multiple high-threshold energy capture layers in the deep crust (see their Fig. 14). This led them to conclude that, collectively, these deeper layers can produce a mass quadrupole capable of matching Eq. (\ref{eq: Torque Balance Quadrupole}), assuming temperature asymmetry at the level $\delta T_{22}/T \sim 1\%$. 

Such conclusions, however, have a couple of caveats that we have already alluded to, and briefly re-summarise: (i) the temperature perturbations that source the displacement of the capture layers are assumed to exist at the percent level \textit{a priori}, and (ii) the crustal composition predicted by HZ90 determined the maximum value of the threshold energy to be $E_{\text{max}}^{\text{HZ90}} = 43.69$ MeV, approximately half the value UCB required to match the torque balance limit for more modest values of the accretion rate $\dot{M}$ and spin frequency $\nu$.

The existence of capture layers in the deep crust of an accreting neutron star is therefore speculative. Aside from the HZ90 model, a newer description of the accreted crust obtained by \citet{Fantina_2018} (hereafter F+18) predicts a maximum threshold energy $E_{\text{max}}^{\text{F+18}} = 69.10$ MeV (See their Tables A1 - A3), still below what UCB required to be sufficient. 

A more recent study by \citet{Gusakov_2020} (hereafter GC20) has even suggested that both the HZ90 and F+18 models are not in fact thermodynamically consistent, since they do not take into account the diffusion of unbound neutrons in the inner crust. When allowing for such diffusion, they found that the EoS of an accreted inner crust should actually be very close to that of cold-catalysed matter (see their Fig. 1). This is true not just for the pressure-density relation, but also for the composition, i.e.\ the run of atomic number $Z$ and mass number $A$. Composition tables for this so-called neutron Hydrostatic and Diffusion (nHD) model can be found in \citet{Potekhin_2023} (Tables 1 and 2), where electron capture takes place only as far as up to neutron drip ($\sim 4 \times 10^{11}$ g cm$^{-3}$), and that $E_{\text{max}}^{\text{GC20}} \approx 25.5$ MeV.

If, for the sake of a simple comparison, we therefore restrict the crustal composition predicted by HZ90 to include just the capture layers preceding neutron drip (the composition of the outer crust is qualitatively similar across all EoS models), then the estimate \eqref{eq: HZ Capture Layers} reduces to just
\begin{equation}\label{eq: HZ Capture Layers 2}
    Q_{\text{tot}} \approx  1.3 \times 10^{36} \, \text{g cm}^{2} \, \mathlarger{\sum}^{4}_i \bigg[ \frac{E_{\text{cap}}^i}{30 \, \text{MeV}} \biggr]^3 \approx 8 \times 10^{35} \, \text{g cm}^{2} \, ,
\end{equation}
and is therefore smaller in magnitude than the torque balance estimate \eqref{eq: Torque Balance Quadrupole}, even before taking into account potential sinking penalties, the possible cancelling out of positive and negative quadrupole moments, and sub-percent temperature asymmetries. 


\subsection{Expected GW emission from capture layer shifts}

The GW torque balance scenario \eqref{eq: Torque Balance Quadrupole} provides a rough upper limit to assess the detectability of CGWs from spinning neutron stars.  It is an upper limit since the well known and well studied interaction between the star and the accretion disk, mediated by the star's magnetic field, may play an important role in determining equilibrium spin frequencies \citep{Ghosh_1977, Ghosh_1979_a, Ghosh_1979_b, White_1997, Andersson_2005}, and may even be sufficient to explain the lack of observed LMXBs rotating faster than $\sim 700$ Hz (e.g. \citealt{Patruno_2012}). 

However, current observations certainly cannot definitively attribute the observed spin-rates of accreting neutron stars to that of the magnetic spin-equilibrium model alone. Indeed, it was noted by \citet{Bhattacharyya_2017} that an external magnetic field $B \sim 10^8$ G (typical of LMXBs) would be insufficient to explain the spin-limit of systems which accrete transiently. Furthermore, a statistical analysis of the spin-distribution of LMXBs by \citet{Patruno_2017} suggests evidence for two distinct sub-populations: the majority of NSs having an average spin frequency $\Bar{\nu} \approx 300$ Hz, and a smaller number of NSs having an average spin frequency $\Bar{\nu} \approx 575$ Hz. A curious aspect of this observation is that if there is indeed an additional braking mechanism in the system, then it must set in sharply once a star reaches a given spin rate. It just so happens that, in the case of a deformed (rotating) NS, the rate of angular momentum loss via GWs scales as a steep (fifth) power of the star’s spin frequency \citep{bild_98}. Another statistical analysis by \citet{Gittins_2019} also found that a qualitatively similar behaviour to the observed distribution of transiently accreting NSs can indeed be obtained from GW spin-equilibrium models. Though, it is worth pointing out that the different GW mechanisms that the authors considered – a permanent quadrupole, unstable r-modes, and thermal mountains (from capture layer shifts) – all produced almost indistinguishable distributions. 


Nevertheless, having estimated the level at which capture layer shifts give rise to GW radiation in the previous section, it is natural to ask whether such mountains are likely to be detected, even if they are not the primary driver of LMXB spin-evolution. 

Armed with the estimate for the mass quadrupole in equation \eqref{eq: UCB Fiducial Q_22}, the amplitude of the corresponding GWs radiated from the NS is \citep{ucb_00}
\begin{equation}\label{eq: GW_amplitude}
    h_0 = \frac{32 \pi}{5} \, \frac{G}{c^4} \, \biggl( \frac{\pi}{3} \biggr)^{1/2} \, \frac{Q_{22} \nu^2}{d} \, ,
\end{equation}
where $G$ and $c$ are the gravitational constant and speed of light, and $d$ is the distance of the given source. 
and $h_0$ is  the GW amplitude \eqref{eq: GW_amplitude} received by an interferometer, assuming optimal orientation of the NS relative to the detector. 

The corresponding signal-to-noise ratio (S/N) as seen in a single detector is given as \citep{Watts_2008}
\begin{equation}
    \biggl(\frac{S}{N} \biggr)^2 = \frac{h_0^2 \, T_{\rm{obs}}}{S_n(f)} \, ,
\end{equation}
where $T_{\rm{obs}}$ is the observation time and $S_n(f)$ is the single-sided power spectral density of the detector. For a given choice of S/N threshold for detectability (which we shall denote as some factor $A$), this leads to a minimum detectable signal amplitude as
\begin{equation}\label{eq: Detector_sensitivity}
    h_{0, \, \rm{det}} = A\sqrt{S_n(f)/T_{\rm{obs}}} \, .
\end{equation}
For targeted searches for CGWs from known pulsars  (where all parameters of the source are known to a sufficiently high accuracy), the factor $A$ is often taken to be $11.4$, corresponding to a specific S/N threshold that permits a single trial false alarm rate of 1\% and false dismissal rate of 10\% \citep{Abbott_2004}. For other types of searches where less is assumed about the signal (e.g. blind all-sky searches), a higher threshold would need to be applied. We will use $A=11.4$ in the analysis here, i.e.\ we assume the phase evolution of the signal is known from electromagnetic observations.

\begin{figure*}
\includegraphics[width=0.9\textwidth]{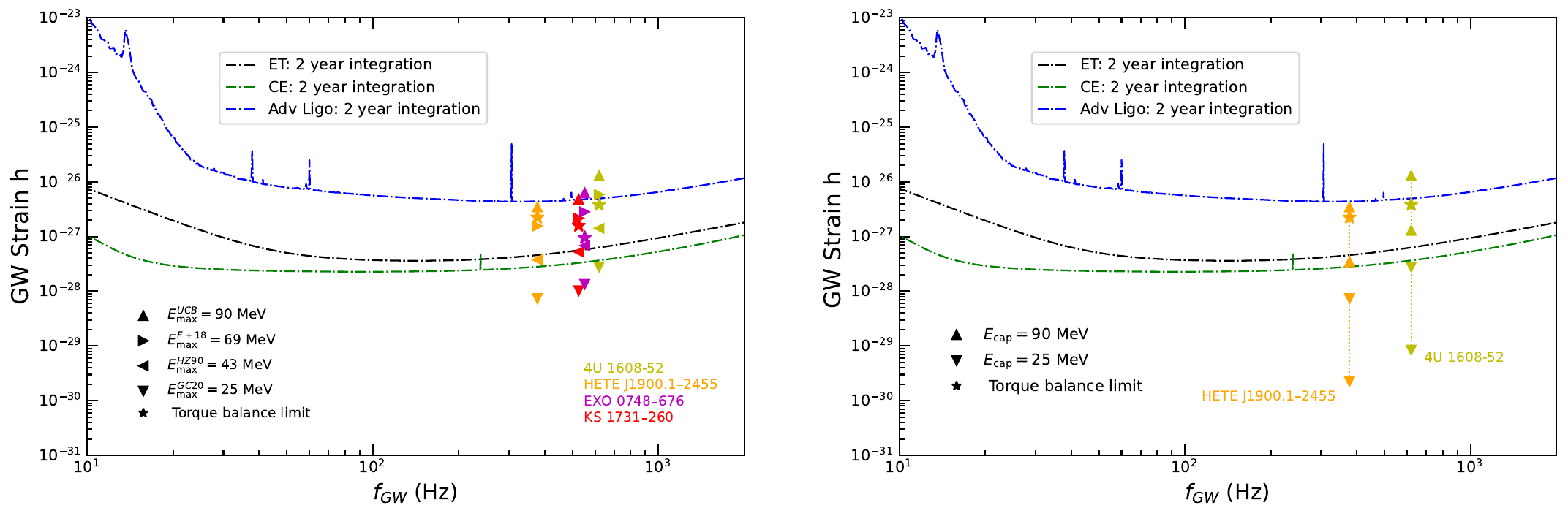} \caption{Plots of signal strength $h_0$ (triangles/stars) and minimum detectable amplitudes $h_{0, \, \rm{det}}$ (curves) for various signal models and GW detectors.  The minimum detectable amplitudes were obtained from equation (\ref{eq: Detector_sensitivity}) with $A=11.4$ and $T_{\rm obs} = 2$ years.  In the left panel we plot the torque balance limit as per equation \eqref{eq: Torque Balance Quadrupole} for the low-mass X-ray binaries listed in Table \ref{tab:LMXBs_table}, and estimate the largest thermal mountain that can be created with an assumed maximum temperature asymmetry $\delta T_{22} / T = 1\%$ for different accreted equations of state (UCB: \citealt{ucb_00}, HZ90: \citealt{hz_90_second}, F+18: \citealt{Fantina_2018}, GC20: \citealt{Gusakov_2020}) using the fiducial estimate \eqref{eq: UCB Fiducial Q_22}. In the right panel we show the effects on mountain sizes which arise from elastic readjustments of the crust to shifting deep (90 MeV) and shallow (25 MeV) capture layers. The bars indicate the
range of uncertainty in $Q_{\rm{fid}}$ (and thus $h_0$) that arise due to `sinking penalties'. 
\label{fig:Strain_plots_capture_layer_shifts}}
\end{figure*}

To contextualise our results, we will compare mountain sizes relative to that of the minimum detectable signal achievable by Advanced LIGO, Cosmic Explorer, and Einstein Telescope, assuming operation at design sensitivities\footnote{Strain sensitivity data was taken from the publicly available LIGO document at: \url{https://dcc.ligo.org/LIGO-T1500293/public}.} with $T_{\rm{obs}} = 2$ years, shown as dash-dotted lines in Figure \ref{fig:Strain_plots_capture_layer_shifts}.


In the left-hand panel of Figure \ref{fig:Strain_plots_capture_layer_shifts} we plot the GW strain as per equation \eqref{eq: GW_amplitude} for the largest thermal mountains that could be formed in a number of known LMXBs (plotted as triangles) by the deepest capture layer (corresponding to $E_{\rm{cap}}^{\rm{max}}$) assuming different accreted equations of state (UCB: \citealt{ucb_00}, HZ90: \citealt{hz_90_second}, F+18: \citealt{Fantina_2018}, GC20: \citealt{Gusakov_2020}) with fixed maximum temperature asymmetry $\delta T_{22} / T = 1\%$.  The quadrupole moment is computed using equation (\ref{eq: UCB Fiducial Q_22}).  The LMXBs we choose to model are given in Table \ref{tab:LMXBs_table}, and are a selection of those considered in \citet{Haskell_2015}.  Specifically, we consider systems which have had an active outburst duration $\delta T_{22}$ longer than two years, and for which the spin period $\nu$, time-averaged accretion rates $\langle \dot{M} \rangle$, and distances $d$ are either known or have been estimated. Only sources which lie above a given noisecurve are considered detectable.

Note that there already exists a direct upper limit on the CGW emission from HETE J1900.1--2455 \citep{lvk_22_AMXPs}, but it is approximately one order of magnitude larger (i.e.\ weaker) than the largest of the amplitudes given for this source in the Figure.


It is clear from Figure \ref{fig:Strain_plots_capture_layer_shifts} that, even with a 2 year observation time, systems with mountains formed from shifting capture layers in the outer crust ($E_{\rm{cap}} < 30$ MeV) not only fail to approach the torque balance limit imposed by equation \eqref{eq: Torque Balance Quadrupole} (marked for each system by stars in the Figure), but also fall below the sensitivity curves of both current \textit{and} next-generation instruments. Even within such an optimistic analysis (i.e.\ ignoring elastic effects and assuming $\delta T_{22} / T = 1\%$), capture layers in the inner crust (in conflict with the inner crust composition computed by \citealt{Gusakov_2020}) must be present if such mountains are to be detected by either Einstein Telescope or Cosmic Explorer.

While capture layers in the deep crust with $E_{\rm{cap}} \geq 90$ MeV exceed the torque balance limit of each of the systems we consider, we reiterate again that the existence of such layers are not supported by any detailed calculations of the accreted equation of state. And although capture layers with $E_{\rm{cap}} \approx 69$ MeV (predicted in the realistic F+18 model) could, in principle, allow for detection of \textit{all} of the systems considered here in next-generation detectors, we also note that temperature asymmetry at the percent level is unlikely to be realised in these systems on account of their low time-averaged accretion rates \citep{oj_20, hj_23}.


The chosen fixed temperature asymmetry $\delta T_{22} / T = 1\%$ is in reality only an upper limit, imposed by the fact that the temperature perturbations which displace the capture layers should result in lateral variations in the flux emanating from the top of the crust. It is thought that if $\delta T_{22} / T \gg 1\%$, the resultant `hot and cold spots' would lead to detectable modulations in quiescent emission \citep{ucb_00}, which so far have yet to be observed.


Regardless, even with such favourable circumstances, as we show in the right-hand panel of Figure \ref{fig:Strain_plots_capture_layer_shifts}, if one does include a `sinking penalty' caused by an elastic readjustment of the crust to capture layers shifts, the results become further pessimistic. Only mountains formed in the very deepest regions of the crust would be large enough for detection; it can be seen that mountains formed in the outer crust now lie \textit{far} below the sensitivity curves. 

Such results certainly provide motivation to explore other methods of sourcing density perturbations in the accreted crust. The thermal contribution to the crustal lattice pressure we shall consider in Section \ref{sect:lattice} is one such method, and is pertinent to this discussion since it does not rely on the existence of capture layers in the deep crust. But first, in Sections \ref{sect:electrons}, and \ref{sect:neutrons}, we look at the simpler cases of thermal contributions to the neutron and proton Fermi gas pressures.




\begin{table*}
\begin{tabular}{l l l l l l l l l l l }
\hline
\hline
Source& &$\nu$ & &  $d$ & & $\langle \dot{M}\rangle$   & &  $\delta T_{22}$& &  Ref.\\

      & & (Hz) & &  (kpc)   & & ($10^{-10}\rm\,M_\odot\,yr^{-1}$) &  &  (days)     & & \\
\hline
HETE J1900.1--2455 && 377   &&  5     &&  8  && 3000&&  \citet{Papitto_2013}\\
EXO 0748--676 $^{(\dag)}$ & & 552     && 5.9     && 3  && 8760& & \citet{Degenaar_2011} \\
4U 1608--52 $^{(\dag)}$ && 620     && 3.6     && 20  && 700&& \citet{Gierliński_2002}\\
KS 1731--260 $^{(\dag)}$ && 526     && 7       && 11    && 4563& & \citet{Narita_2001}\\
\hline
\end{tabular}

\caption{Estimated spin frequency $\nu$, distance $d$, and time-averaged accretion rate $\langle \dot{M} \rangle$ of a number of low-mass X-ray binaries with observed outburst duration's $\delta T_{22}$ longer than 2 years. Sources marked $^{(\dag)}$ are nuclear powered pulsars, while HETE J1900.1--2455 is an accreting millisecond pulsar. Note that this is an adapted reproduction of Table 1 from \citet{Haskell_2015}.}

\label{tab:LMXBs_table}
\end{table*}

\section{Thermal corrections to the electron gas pressure}  \label{sect:electrons}

To a good approximation, when considering the structure of a mature NS, one can use a zero-temperature EoS, i.e.\ neglect thermal effects completely.  However, there will be some small thermal correction for all reasonable models of dense matter, tending to increase the pressure with respect to the zero-temperature approximation.  It is this small thermal correction, coupled to a non-axisymmetry in the temperature distribution, that would generate the elasto-thermal mountain that we seek to model.  To gain some preliminary insight into thermal effects, we first look at the analytically-tractable case of a Fermi gas.


The pressure in the outer part of a NS crust can be approximated as coming from a Fermi gas of relativistic electrons. The Fermi temperature $T_{\rm F}$ gives an estimate of the temperature at which thermal effects become important (\citet{ll_69}, Section 61):
\be
T_{\rm F} =
(3\pi^2)^{1/3} \frac{\hbar c}{k_{\rm B}}  \left(   \frac{\rho}{\mu_{\rm e} m_{\rm u}}  \right)^{1/3} 
=
4.15 \, \times \, 10^{11} \, {\rm K \,} \rho_{12}^{1/3}  \left(\frac{3}{\mu_{\rm e}}\right)^{1/3} .
\ee
where $\rho$ is the mass density, $\mu_{\rm e}$ the mean molecular weight (i.e.\ average number of baryons per electron), and $m_u = 1.66 \times 10^{-24}$\,g the atomic mass unit.  The relevant dimensionless parameter that describes the importance of finite-temperature effects is the ratio of the actual temperature $T$ to $T_{\rm F}$:
\be
\frac{T}{T_{\rm F}} = 
T \frac{1}{(3\pi^2)^{1/3}} \frac{k_{\rm B}} {\hbar c} \left(   \frac{\mu_{\rm e} m_{\rm u}}{\rho}  \right)^{1/3} 
=
2.42 \, \times \, 10^{-4} \, T_8 
\left(\frac{\mu_{\rm e}}{3}\right)^{1/3} 
\frac{1}{\rho_{12}^{1/3}}  .
\ee
As expected, for the temperatures of $\sim 10^8$\,K typical of NSs in LMXBs, the effects of temperature will be small.

In the limit of small $T/T_{\rm F}$, the total pressure can be divided into $P_0$, the zero-temperature piece, and an (analytically-calculable) thermal correction $P_{\rm therm}$.  Using results from \citet{ll_69}, sections 61.4--61.6, we obtain
\be
P = P_0 + P_{\rm therm} = P_0 \left[ 1 + 2 \pi^2 \left(\frac{T}{T_{\rm F}}\right)^2 \right ] .
\ee
Note that the thermal correction to the pressure is \emph{quadratic} in the parameter $T / T_{\rm F}$.

The fractional increase in pressure due to finite-temperature effects is then
\be
\frac{P_{\rm therm} }{P_0} = 2 \pi^2 \left(\frac{T}{T_{\rm F}}\right)^2 
= 
1.15 \times 10^{-6} \, T_8^2 \left(\frac{\mu_{\rm e}}{3}\right)^{2/3} \frac{1}{\rho_{12}^{2/3}} .
\ee

\begin{figure*} 
\includegraphics[width=0.8\textwidth,height=6cm]{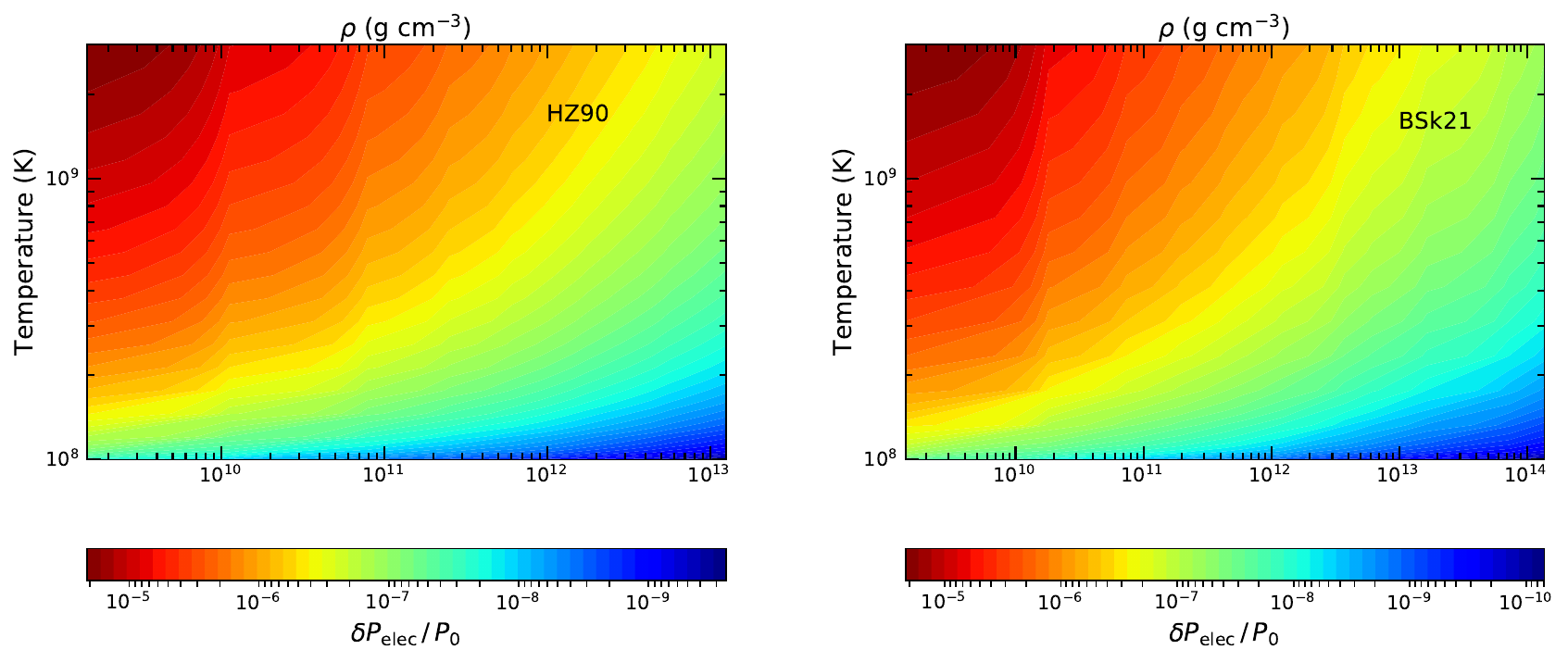} \caption{Fractional perturbation in pressure due to relativistic electrons for the \citet{hz_90_first, hz_90_second} (\textit{left}) and BSk21 (\citealt{Fantina_2018, Fantina_2022}; \textit{right}) equations of state, as a function of density and temperature, for a temperature perturbation $\delta T_{22} / T = 1\%$, computed using equation (\ref{eq:delta_P_over_P_e}). \label{fig:delta_P_electron_over_P_new}}
\end{figure*}

Most relevant for mountains is the perturbation in pressure caused by a perturbation $\delta T_{22}$ in $T$:
\be
\label{eq:delta_P_over_P_e}
\frac{\delta P_{\rm therm} }{P_0} = 4 \pi^2 \left(\frac{T}{T_{\rm F}}\right)^2 \left(\frac{\delta T_{22}}{T}\right)
= 
2.29 \times 10^{-8} \, \left(\frac{\mu_{\rm e}}{3}\right)^{2/3} \frac{1}{\rho_{12}^{2/3}} 
T_8^2   \left(\frac{\delta T_{22} / T}{1 \%}\right)   ,
\ee
where we have parameterised the temperature perturbation as a $1\%$ change to the unperturbed value. We plot this fractional change in Figure \ref{fig:delta_P_electron_over_P_new}, for an assumed uniform fractional temperature perturbation of $\delta T_{22} / T = 1\%$.  The fractional perturbation in pressure can be seen to vary from $\sim 10^{-5}$ in the outer crust, down to $\sim 10^{-10}$ near the crust-core transition. The implications of these results will be discussed in Section \ref{sect:lattice}, where we shall compare the fractional perturbation in pressure due to each of the electrons, neutrons, and crustal lattice.

\section{Thermal corrections to the neutron gas pressure}  \label{sect:neutrons}

The pressure in the inner part of the NS crust can be approximated as coming from a Fermi gas of non-relativistic neutrons.  The Fermi temperature is given in terms of the \emph{free} neutron number density
$n_{\rm i}$, which can be written in terms of the free neutron fraction $X_{\rm n}$
\be
n_{\rm i} = \frac{\rho}{m_{\rm u}} X_{\rm n} .
\ee
From \citet{ll_69}, section 57.3:
\be
T_{\rm F} = (3\pi^2)^{2/3} \frac{\hbar^2}{2 m_{\rm u} k_{\rm B} }  n_{\rm i}^{2/3} 
=
1.66 \times 10^{10} {\, \rm K \,} (\rho_{12} X_{\rm n})^{2/3} . 
\ee
The dimensionless parameter giving the importance of finite-temperature corrections is then
\be
\frac{T}{T_{\rm F}} = 6.09 \times 10^{-3} \frac{T_8}{ (\rho_{12} X_{\rm n})^{2/3}} ,
\ee
again indicating that thermal corrections will be small for NSs in LMXBs.

In the case of small $T / T_{\rm F}$, the pressure can again be divided into zero-temperature and thermal pieces.  Using results for sections 56--58 of \citet{ll_69} we obtain
\be
P = P_0 + P_{\rm therm} = P_0 \left[ 1 + \frac{5\pi^2}{8} \left(\frac{T}{T_{\rm F}}\right)^2 \right] ,
\ee
again quadratic in $T/T_{\rm F}$.

The increase in pressure due to finite-temperature effects is then
\be
\frac{P_{\rm therm} }{P_0} = \frac{5\pi^2}{8} \left(\frac{T}{T_{\rm F}}\right)^2 
= 
2.29 \times 10^{-4} \frac{T_8^2}{(\rho_{12} X_{\rm n})^{4/3} } .
\ee

Most relevant for mountains is the perturbation in pressure caused by a perturbation $\delta T_{22}$ in $T$:
\be
\label{eq:delta_P_over_P_n}
\frac{\delta P_{\rm therm} }{P_0} = \frac{5\pi^2}{4} \left(\frac{T}{T_{\rm F}}\right)^2 \left(\frac{\delta T_{22}}{T}\right)
= 
4.50 \times 10^{-6} \frac{T_8^2}{(\rho_{12} X_{\rm n})^{4/3} } \left(\frac{\delta T_{22} / T}{1 \%}\right)  .
\ee

\begin{figure*} 
\includegraphics[width=0.8\textwidth,height=6cm]{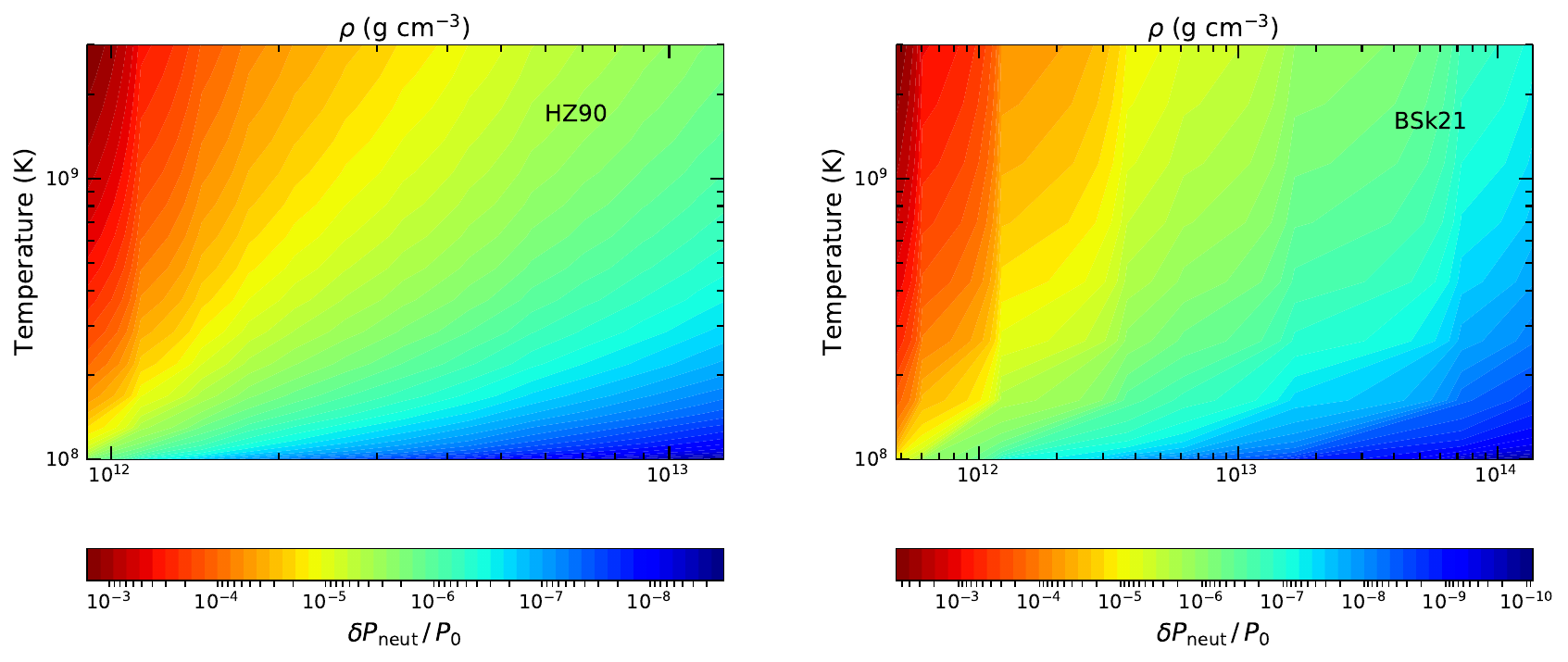} \caption{Fractional perturbation in pressure due to non-relativistic neutrons for the \citet{hz_90_first, hz_90_second} (\textit{left}) and BSk21 (\citealt{Fantina_2018, Fantina_2022}; \textit{right}) equations of state, as a function of density and temperature, for a temperature perturbation $\delta T_{22} / T = 1\%$, computed using equation (\ref{eq:delta_P_over_P_n}).  \label{fig:delta_P_neutron_over_P_new}}
\end{figure*}

Like before, we plot this fractional change in Figure \ref{fig:delta_P_neutron_over_P_new}, again for an assumed uniform fractional temperature perturbation of $\delta T_{22} / T = 1\%$. We use a lower density equal to that of neutron drip ($6.11 \times 10^{11}$ g cm$^{-3}$ for HZ90 and $4.38 \times 10^{11}$ g cm$^{-3}$ for BSk21), as this effect requires the existence of free neutrons. The magnitude of the fractional perturbation can be seen to range from $\sim 10^{-3}$ at the onset of neutron drip, down to $\sim 10^{-10}$ near the crust-core transition at low temperature.

\section{Thermal corrections to the lattice pressure}  \label{sect:lattice}

The thermal corrections to the pressures above are small, but given that the level of non-axisymmetry required to produce astrophysically interesting and potentially detectable levels of CGW emission are also small, they are nevertheless of interest.  However, the results apply to (Fermi) \emph{gases}, and it therefore seems plausible that any such thermally-induced perturbations in density would simply convect away.  For the neutrons, this seems almost inevitable.  The electrons are electrically charged, and so would feel the effects of the large scale stellar magnetic field; it is possible that the magnetic field may play a stabilising role, giving rise to a \emph{thermo-magnetic} mountain.

We will not pursue this idea further here.  Rather, we will turn our attention to a piece of the thermal crustal pressure that is necessarily firmly tied to the elastic phase.  Specifically, we will look at the \emph{crustal lattice pressure}.  This is the pressure that is produced directly by the ionic lattice itself, through interactions of the ions with other ions, or the background sea of electrons.  The crustal lattice pressure is small compared to the total pressure (which can be approximated by the Fermi gas analysis described above), and its thermal correction will be smaller still.  However, the fact that it is tied to the lattice makes it potentially interesting from the point of view of mountain building.

The thermodynamic properties of the neutron star crust are described in detail in \citet{hpy_07}, whose treatment and notation we largely follow here.

There are two dimensionless numbers that are important when discussing the thermal properties of crystals.  The first is the \emph{ion Coulomb parameter} $\Gamma$, the ratio of the ions' electrostatic potential energy to their thermal energy (\citealt{hpy_07}, equation  (2.22)):
\be
\Gamma \equiv \frac{Z^2 e^2}{a_{\rm i} k_{\rm B} T} ,
\ee
where $Z$ is the atomic number of the ions and $a_{\rm i}$ the ion sphere radius, defined by (\citealt{hpy_07}, equation (2.23)) as
\be
a_{\rm i} = \left(\frac{3}{4 \pi n_{\rm i}}\right)^{1/3} ,
\ee
where $n_{\rm i}$ is the number density of ions.  The Coulomb parameter is useful in two ways.  Firstly, it is found that one component plasmas melt for $\Gamma \lesssim 175$.  Secondly, it measures the importance of so-called \emph{anharmonic corrections} to the low-temperature harmonic treatment of ionic oscillations.  Basically, for sufficiently high $T$ (i.e.\ sufficiently small $\Gamma$), the ions' oscillations are sufficiently large that one needs to go beyond a quadratic form for the potential in which they oscillate; see Section 2.3.3e of \citet{hpy_07}.

The second important dimensionless number is the ratio of the temperature $T$ to the \emph{plasma ion temperature} $T_{\rm pi}$.   This is defined in terms of the \emph{plasma ion frequency} $\omega_{\rm pi}$ (\citealt{hpy_07}, equation (2.30)):
\be
\label{eq:omega_pi}
\omega_{\rm pi}^2 = 4 \pi e^2 n_{\rm i} \left\langle \frac{Z^2}{m_i} \right\rangle ,
\ee
where the ion density $n_{\rm i}$ is given by 
\be
\label{eq:n_N}
n_{\rm i}  = \frac{\rho (1 - X_{\rm n})}{m_{\rm u} A} ,
\ee
and the angle brackets denotes an average over the atomic numbers $Z$ and masses $m_i$ of the ions.

From equation (2.29) of \citet{hpy_07} we have \be
T_{\rm pi} = \frac{\hbar \omega_{\rm pi}}{k_{\rm B}} ,
\ee
and so
\be
\frac{T}{T_{\rm pi}} = \frac{k_{\rm B} T}{\hbar \omega_{\rm pi}} .
\ee
For $T / T_{\rm pi} \ll 1$ we are in the \emph{quantum regime}, while for $T / T_{\rm pi} \gg 1$ we are in the \emph{classical regime}.

The description of the thermal properties of a crystal can then be divided into a low temperature harmonic regime, a high temperature anharmonic region, and, simultaneously and separately, into a low temperature quantum regime and a high temperature classical regime.  In what follows we will assume we are always in the harmonic regime, i.e. at temperatures well below the melting temperature.  This is a good approximation for NSs in LMXBs, for all but the lowest density parts of the crust.

In the quantum regime, the thermal  piece of the pressure can be obtained by differentiation of the  lattice's free energy (equation (2.109) of \citealt{hpy_07}) with respect to volume, to give
\be
P_{\rm th}(T \ll T_{\rm pi})  = \frac{\zeta n_{\rm i}}{8} \left(\frac{T}{T_{\rm pi}}\right)^4  k_{\rm B} T_{\rm pi} .
\ee
where $\zeta$ is a constant of order unity. 

For temperatures $T \gg T_{\rm pi}$, the thermal energy of the lattice can be treated in a classical way.  The lattice's free energy is given in equation (2.106) of \citet{hpy_07}, which can be differentiated with respect to volume to give the thermal pressure:
\be
\label{eq:P_th_classical}
P_{\rm th}(T \gg T_{\rm pi})  = \frac{3}{2} n_{\rm i} k_{\rm B} T .
\ee

\begin{figure*} 
\includegraphics[width=0.8\textwidth,height=6cm]{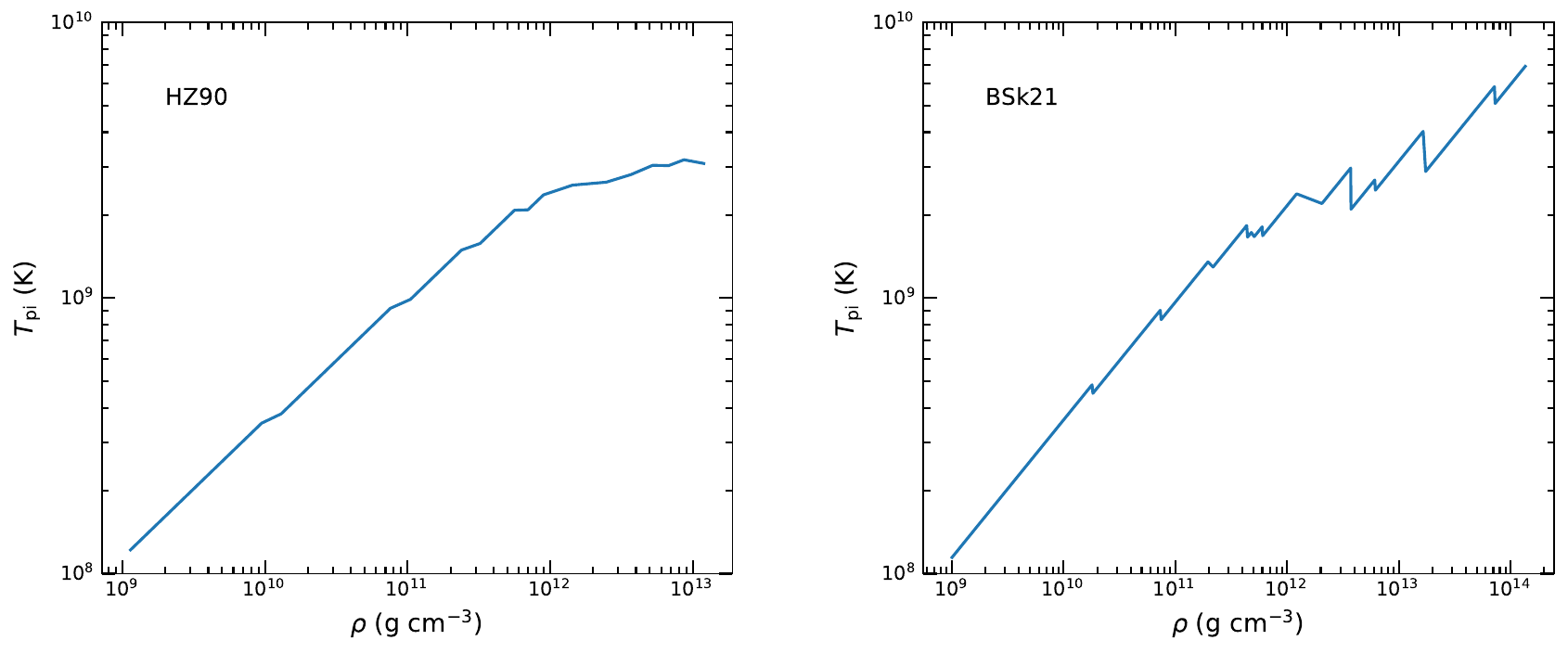} \caption{Plasma ion temperature $T_{\rm pi}$ for the \citet{hz_90_first, hz_90_second}  (\textit{left}) and BSk21 (\citealt{Fantina_2018, Fantina_2022}; \textit{right}) equations of state. \label{fig:T_pi_new}}
\end{figure*}

To decide which regime is applicable, we plot the plasma ion temperature for both the \cite{hz_90_first, hz_90_second} EoS  and the BSk21 \citet{Fantina_2018, Fantina_2022} EoS in Figure \ref{fig:T_pi_new}.  A typical NS in an LMXB has a temperature of a few times $10^8$\,K, so that $T / T_{\rm pi} \sim 0.1$ over much of the crust, for both equations of state. This indicates that we are in the quantum regime, but not deeply so.

For this reason, we will not use results from the limiting cases of $T / T_{\rm pi} \ll 1$ or $T / T_{\rm pi} \gg 1$  to compute the thermal pressure.  Instead, we will use the results of \citet{bpy_01}, as they are valid for arbitrary $T / T_{\rm pi}$.  Their results are reproduced on page 83 of \citet{hpy_07}, with slightly different notation.

\citet{bpy_01} write the thermal part of the free energy, $F_{\rm th}$, in terms of the reduced thermal free energy $f_{\rm th}$:
\be
F_{\rm th} = f_{\rm th} n_{\rm i} k_{\rm B} T ,
\ee
where we have converted from their notation of $N$ to $n_{\rm i}$.  They give a fitting function for $f_{\rm th}$ in terms of a parameter $\theta$:
\be
\label{eq:f_th_and_theta}
f_{\rm th} = f_{\rm th}(\theta), \hspace{10mm} \theta \equiv \frac{\hbar \omega_{\rm pi}}{k_{\rm B} T} = \frac{T_{\rm pi}}{T} ,
\ee
with the (rather elaborate) fitting function given in their equations (13) and (14).

The pressure is related to the free energy by (\citealt{hpy_07}, equation (2.36)):
\be
\label{eq:P_dFbdV}
P = - \left.\frac{\partial F}{\partial V}\right|_{n_{\rm i} , T} .
\ee
Making use of this, exploiting the fact that the volume dependence of $F$ is contained within the dependence of $\theta$ on $n_{\rm i}$, we obtain
\be
\label{eq:P_th}
P_{\rm th} =  \frac{1}{2} \hbar \omega_{\rm pi} n_{\rm i}  \frac{df_{\rm th}}{d\theta} .
\ee
This can be used to compute the thermal pressure for a given EoS, making use of equations (\ref{eq:n_N}), (\ref{eq:omega_pi}) and (\ref{eq:f_th_and_theta}), with $f_{\rm th}'(\theta)$ easily obtainable by differentiating the fitting formula given in \citet{bpy_01}, using their equations (13) and (14). 



\begin{figure*} 
\includegraphics[width=0.8\textwidth,height=6cm]{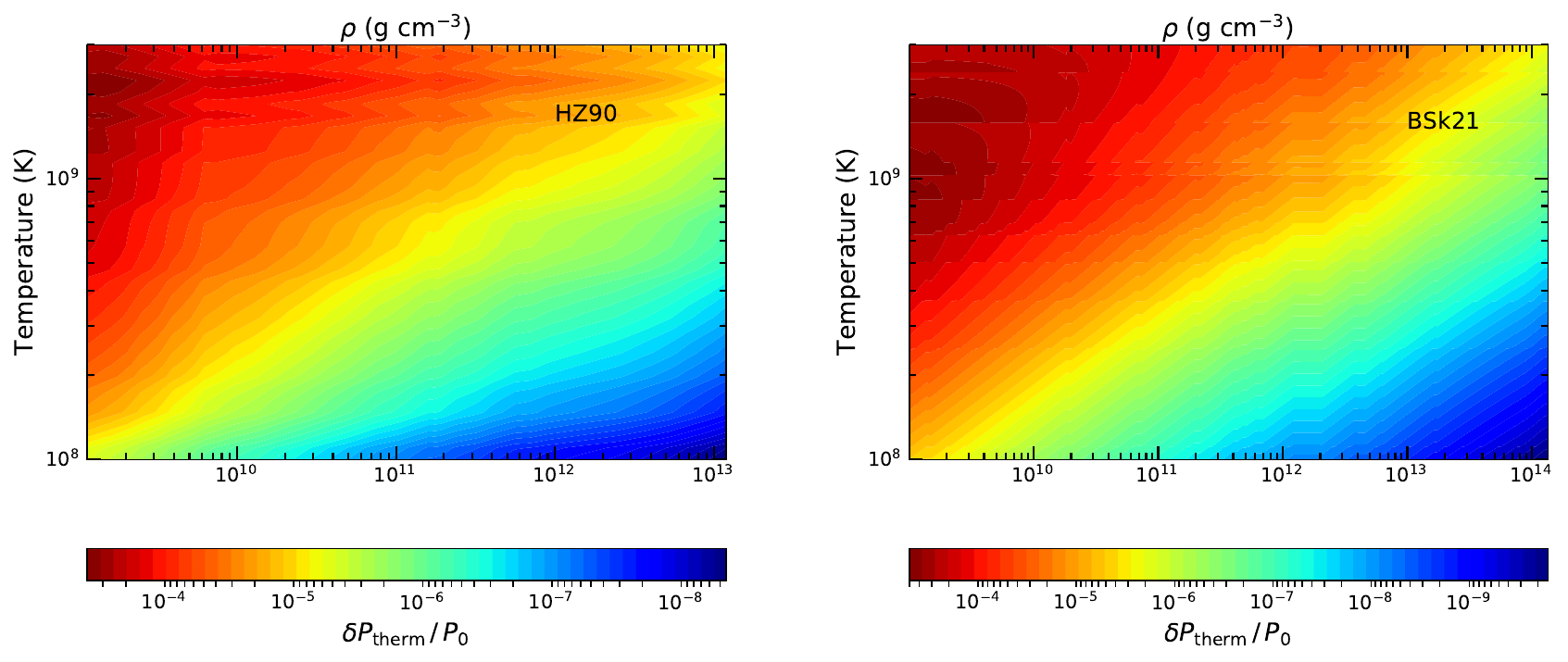} \caption{Fractional perturbation in pressure due to the crustal lattice for the \citet{hz_90_first, hz_90_second} (\textit{left}) and BSk21 (\citealt{Fantina_2018, Fantina_2022}; \textit{right}) equations of state, as a function of density and temperature, as computed using the formalism described in Section \ref{sect:lattice}, for a temperature perturbation $\delta T_{22} / T = 1\%$.  \label{fig:delta_P_thermal_over_P_new}}
\end{figure*}

For mountain building, it is more insightful to consider the perturbation in the thermal pressure, $\delta P_{\rm th}$, caused by a temperature perturbation $\delta T_{22}$, rather than $P_{\rm th}$ itself.  From equations (\ref{eq:f_th_and_theta}) and (\ref{eq:P_th}) we obtain 
\be
\frac{d P_{\rm th}}{dT} = - \frac{1}{2} \hbar \omega_{\rm pi} n_{\rm i} \frac{1}{T} \theta
\frac{d^2 f_{\rm th}(\theta)}{d\theta^2} .
\ee
The derivative of $f^{''}_{\rm th}(\theta)$ can easily be evaluated from the fits given in \citet{bpy_01}.  This derivative can then be inserted into 
\be
\label{eq:source_term}
\frac{\delta P_{\rm th}}{P_0} = \frac{T}{P_0} \frac{dP_{\rm th}}{dT} \left(\frac{\delta T_{22}}{T}\right) 
\ee
to give the fractional change in pressure caused by a fractional temperature perturbation $\delta T_{22} / T$.  Note that is it essentially this quantity that acts as the source term in the full perturbative analysis of the elastic equilibrium problem solved in \citet{ucb_00}; see their equation (35) (but, of course, their perturbations were sourced by moving electron capture layers, not the thermal piece of the crustal EoS considered here.)  A plot of this fractional change is given in Figure \ref{fig:delta_P_thermal_over_P_new}, for an assumed uniform fractional temperature perturbation of $\delta T_{22} / T = 1\%$.  The fractional perturbation in pressure can be seen to vary from $\sim 10^{-4}$ in the outer crust, down to $\sim 10^{-6}$ in the deep crust.

The importance of considering the pressure perturbation $\delta P_{\rm th}$ stems from the fact that the ratio $P_{\rm th} / P_0$ would describe only the change in \textit{spherical} structure induced by the thermal component of the pressure relative to a zero-temperature star. As such, this has no direct observational significance. In contrast, the (relatively small) pressure perturbations of Figure \ref{fig:delta_P_thermal_over_P_new} refer to \textit{non-spherical} changes, and thus are potentially observable via the gravitational waves the associated mass quadrupole would produce (assuming the star is rotating).

Before comparing the results we have obtained for pressure variations due to thermal lattice pressure with those of UCB for capture layers shifts (see Section \ref{sect:comparison}), we will first make a quick comparison between the electron gas, neutron gas, and lattice pressure results given above.  UCB found that much of their quadrupole was generated for densities of a few times $10^{12}$ g\,cm$^{-3}$.  A typical LMXB temperature is a few times $10^8$ K in this density range (see e.g.\ \citealt{hj_23}).  If we look at this portion of the density-temperature parameter space of Figures \ref{fig:delta_P_electron_over_P_new},  \ref{fig:delta_P_neutron_over_P_new} and
\ref{fig:delta_P_thermal_over_P_new}, we see that the fractional pressures perturbations for electrons, neutrons, and the crustal lattice, are $\sim 10^{-7}$, $\sim 10^{-6}$, and $\sim 10^{-6}$, respectively.  It seems that the thermal component of crustal lattice pressure is competitive with the thermal fluctuations in the Fermi gas contributions, as well as having the advantage of being firmly tied in place by the crustal lattice itself.


\section{Comparison between capture layer shifts and thermal crustal pressure}  \label{sect:comparison}

Having examined the effect of temperature on crustal lattice pressure, it is natural to try and compare its importance with the effective temperature dependence of the EoS that comes through the capture layer shifts considered in \citet{ucb_00}.  Rather than repeating their entire calculation for our neutron star model, we will make a rough comparison using their published results. 

Specifically, we will, for a given temperature perturbation $\delta T_{22}$, compare the sizes of the corresponding pressure perturbations $\delta P$, for our crustal lattice pressure perturbations and UCB's shifting capture layers.  The justification for using this simple method of comparison lies in the fact that the relevant set of ODEs (equations 43(a)--43(d) of UCB) are linear in both the crust's elastic displacement and in the source term itself, implying that larger source terms automatically produce larger elastic responses.  Of course, the exact radial profile of the perturbation would differ in the two cases, the effect of which can only be quantified with a full numerical solution (which we will present in a subsequent work).

In particular, the source terms in UCB are very sharply peaked at the capture layers that divide shells of different atomic number, whereas our source terms as smooth, being directly obtained from the star's thermal profile, as per our equation (\ref{eq:source_term}).  In making our rough comparisons below, we are implicitly assuming that this sharpness does not introduce any qualitatively different features in the elastic response of the crust and the resultant density perturbations, or any sort of ``screening'' effect where perturbations at different densities nearly cancel one another.  Such a screening effect has been seen in the tidal responses of two layer quark stars \citep{letal_19}, but there is no reason to expect it to apply in our case.

UCB give plots showing results for the particular case of a star with a normal (i.e. not superfluid) core, accreting at a (rather high) rate of $0.5 \dot M_{\rm Edd}$, half the Eddington accretion limit.  Their temperature asymmetry is generated by an asymmetry in the local nuclear heating rate, but this is not of importance for the purposes of this comparison, only the temperature perturbation itself is needed to compute the corresponding pressure perturbation.    This temperature perturbation is given in the upper left panel of their Figure 8, and can be seen to be at the few percent level, specifically $\delta T_{22} / T \sim 3\%$ over the density range where their capture layers occur.  

The corresponding fractional perturbations in pressure, as defined by our equation (\ref{eq:source_term}) for this $\sim 3\%$ temperature perturbation, are then given in the lower panels of UCB's Figures 10-12.  In their Figure 10 they give results for a shallow capture layer with capture energy $E_{\rm cap} = 23$ MeV.  In their Figure 11 they give results for a deeper capture layer with $E_{\rm cap} = 42$ MeV, corresponding to one of the deepest capture layers in the computation of \citet{hz_90_second}.  In their Figure 12 they gives results for a capture layer with $E_{\rm cap} = 95$ MeV.  As we noted in Section  \ref{sect:capture_layers}, no such capture layer was found in the computations of \citet{hz_90_second} or \citet{Fantina_2018}.  Given this, we will focus our comparison on the second of these capture layers.

For this $E_{\rm cap} = 42$ MeV capture layer, Table 2 of \citet{hz_90_second} shows that $\rho \approx 9.0 \times 10^{12}$ g cm$^{-3}$ and $P \approx 1.2 \times 10^{31}$ erg cm$^{-3}$.  For the particular accreting star that UCB consider, their Figure 7 indicates that $T \approx 6 \times 10^8$ K.  Note that, from the left panel of our Figure \ref{fig:delta_P_thermal_over_P_new}, we have $\delta P_{\rm th} / P_0 \approx 3 \times 10^{-6}$ for this combination of density and temperature and temperature perturbation $\delta T_{22} / T \approx 3 \%$, assuming (for consistency) the EoS of \citet{hz_90_first, hz_90_second}.  In contrast, the pressure perturbation of UCB, from the bottom panel of their Figure 11, peaks at around $\delta P_{\rm th} / P \sim 10^{-2}$.  This indicates that within their capture layer, the thermal component of the pressure is very important: a percent-level temperature variation is resulting in a percent-level pressure perturbation.  

However, the function is extremely sharply peaked in radius, with a full-width half maximum $\sim 10^2$ cm.  This sharpness stands in contrast to the smooth variation in pressure with radius that holds for the crustal lattice pressure perturbations; observe the smooth variation of $\delta P$ with density in Figure \ref{fig:delta_P_thermal_over_P_new}.  Physically, this is related to the fact that for UCB, the pressure perturbation is coming from the change in the (spatially thin) capture layer, whereas for us it comes from the crustal lattice pressure, which varies smoothly over the crust.  

Given this, in comparing UCB's pressure perturbation with ours, we first carry out an average with respect to radius, the radius being the relevant independent variable on the set of linear ODEs that determine the elastic response of the star (again, see equations 43(a)--43(d) of UCB).  Averaging over the crustal thickness of $\sim 10^5$ cm, the radially averaged fractional pressure perturbation is $\langle \delta P_{\rm th} / P \rangle \sim 10^{-5}$.  Allowing for the existence of $\sim 10$ such capture layers gives $\langle \delta P_{\rm th} / P \rangle \sim 10^{-4}$.

We therefore see that the the capture layer pressure perturbations used in UCB do indeed dominate over crustal lattice pressure perturbation, but only by a factor of order $\sim 10^{-4} / (3 \times 10^{-6}) \sim 30$.  If instead one were to use the more recently computed results of \citet{Potekhin_2023}, then there would be fewer capture layers to sum over, and their threshold energies would be lower, around $E_{\rm cap} \sim 20$ MeV.  UCB found that such (relatively) low energy captures produce mass quadrupoles that are more than an order of magnitude smaller than the $E_{\rm cap} = 42.4$ MeV capture layer; compare their Figures 10 and 11.  On this basis, it seems that a more detailed exploration of the mass quadrupoles that can be built with the crustal lattice pressure seems worthwhile.

As a caveat, we remind the reader that in a real neutron star, the pure $l=m=2$ deformations considered here will exist alongside other multipole contributions, dependent upon the history of the crust since formation.  The entire history of strain, fracture and plastic flow will be relevant.  These extra contributions to the strain will add, and it is the sum of these that will ultimately be limited by the finite strength of the crust.  See the material surrounding equation (34) of UCB for more discussion of this.

\section{Summary and Discussion}  \label{sect:discussion}

Gravitational waves from thermoelastic mountains may play a role in determining the spin frequencies of neutron stars in LMXBs \citep{bild_98}.  The leading candidate for producing such deformations relies on the existence of electron capture layers with high threshold energies, deep in the crust \citep{ucb_00}.  As we described in Section \ref{sect:capture_layers}, modern EoS calculations do not provide support for the existence of such deep layers \citep{Fantina_2018, Fantina_2022}, with a recent study even casting doubt over the existence of there being \emph{any} capture layers at densities above neutron drip \citep{Gusakov_2020, Potekhin_2023}.  

This motivated us to look elsewhere for temperature-dependencies in the neutron star EoS.  In Section \ref{sect:electrons} we considered the finite temperature effects on the relativistic electron gas that provides the pressure support in the outer crust, while  in Section \ref{sect:neutrons} we did the same for the non-relativistic neutron gas that provides pressure support in the inner crust.  We found that 
these thermal effects can  provide small but potentially interesting pressure perturbations.  However, being perturbations in a fluid, neither of these seem good candidates for producing a stable mass asymmetry, as would be required for long-lived gravitational wave emission.

In Section \ref{sect:lattice} we therefore turned out attention to the thermal component of the pressure supplied by the crustal lattice itself.  We showed that while small, thermal effects in this pressure may be relevant, making a crude comparison with the capture layer shifts in Section \ref{sect:comparison}.  

It should also be noted that whereas the capture layer shift mechanism applies only in accreting systems, the crustal pressure mechanism is potentially relevant in \emph{all} neutron stars, both accreting and isolated.  All that is required in a sufficiently large temperature asymmetry and, for significant gravitational wave emission, a sufficiently high spin frequency.

We therefore advance such crustal lattice pressure asymmetries as worthy of further investigation.  We are carrying out such a study ourselves, with the star's magnetic field providing the source of the temperature asymmetry, building on the work presented in \citet{oj_20} and \citet{hj_23}.  We will present this in a separate study.

\section*{Acknowledgements}

We wish to thank Anthea Fantina and Nicolas Chamel for useful discussions, and for sharing data on the pressure-density relations for the BSk19-21 EoSs, as well as Ian Hawke for providing assistance with the construction of our numerical code. TH acknowledges support from the Science and Technology Facilities Council (STFC) through Grant No. ST/T5064121/1. DIJ acknowledges support from the STFC via grant No. ST/R00045X/1.

\section*{Data Availability}

No new data were generated or analysed in support of this research.



\bibliographystyle{mnras}
\bibliography{bibliography} 








\bsp	
\label{lastpage}
\end{document}